\documentclass[10pt]{iopart}
\usepackage{epsfig}

\newcommand{\be}{\begin{equation}}
\newcommand{\ee}{\end{equation}}
\newcommand{\pizero}{\mbox{($\pi$,0) }}
\newcommand{\pihalf}{\mbox{($\pi/2$,$\pi/2$) }}
\begin{document}


\jl{3}
\letter{Quantum renormalization of high energy excitations in the 2D Heisenberg antiferromagnet}
\author{O F Sylju{\aa}sen{\dag}  and H M R{\o}nnow\ddag}
\address{\dag\ NORDITA, Blegdamsvej 17, DK-2100 Copenhagen {\O}, Denmark}
\address{\ddag\ Condensed Matter Physics and Chemistry Department,
  Ris\o{} National Laboratory, DK-4000 Roskilde, Denmark}

\begin{abstract}
We find using Monte Carlo simulations of the spin-1/2 2D square 
lattice nearest
neighbour quantum Heisenberg antiferromagnet that the high energy peak locations at
\pizero and \pihalf differ by about 6\%, \pihalf  being the
highest. This is a deviation from linear spin wave theory which
predicts equal magnon energies at these points.
\end{abstract}


The simplest model describing quantum antiferromagnets is the nearest
neighbour quantum Heisenberg model. 
Among the class of materials which
to a good accuracy can be described by this model are the undoped
high-temperature superconductors where the strongly interacting
spins are located on a two-dimensional square lattice.
Although simple to formulate, the Heisenberg model is not exactly
solvable in dimensions greater than one, and approximations or
numerical calculations are needed to compare the predictions of the
Heisenberg model to experiments.

For the 2D $S=1/2$ Heisenberg antiferromagnet on a square lattice,
the time independent properties are well 
understood \cite{CHN,cuccoli97,beard98}, and 
measurements of for instance the correlation length 
\cite{Greven,birgeneau99,RMH,carretta00} agree
very well with the theoretical predictions.
The situation concerning the dynamics is more unclear,
and there is a need for definite predictions to which experiments can
be compared.
In particular, recent neutron scattering measurements on
Cu(DCOO)$_2\cdot$4D$_2$O \cite{Ronnow} and La$_2$CuO$_4$ \cite{Aeppli}
directly probe the magnon dispersion between the two points \pihalf
and \pizero on the Brillouin zone boundary.
These two materials, which are both considered to be physical
realizations of the 2D $S=1/2$ Heisenberg antiferromagnet, 
show respectively a 6\% decrease and
a 13\% increase in the magnon energy between \pihalf and ($\pi$,0).
These results are in contrast to the linear
spin-wave approximation of the 2D Heisenberg model, which predicts equal
magnon-energies at these points. 
While these deviations from linear spin-wave theory could be due to
additional terms in the Hamiltonian describing each of the materials, 
it is also possible that there are corrections to linear spin-wave theory.
In this Letter we aim at clarifying the predictions for the
$S=1/2$ Heisenberg antiferromagnet on a square lattice at high
energies, in particular at the special
points \pizero and \pihalf in the Brillouin zone. 

The linear spin-wave approximation which is the zeroth term in an
expansion in the parameter $1/S$ gives the magnon dispersion
\be \label{LSW}
 \omega_k = 4JS \sqrt{1- \gamma_k^2},
\ee
where $\gamma_k = (\cos{k_x}+\cos{k_y})/2$. The wave-numbers are
measured in units of the lattice spacing. Note that $\gamma$ is zero
both at \pizero and ($\pi/2$,$\pi/2$). 

The effect of the first order correction to linear
spin-wave theory is to renormalize uniformly the magnon dispersion by a factor
$Z=1.158$. While there is a question as to what extent one can trust the spin-wave expansion for $S=1/2$ it has been argued from Monte Carlo measurements \cite{Chen} that the {\em only}
correction to linear spin-wave theory is such a uniform renormalization of
the spectrum. Based on measurements at low energies this renormalization constant is found to be $Z=1.183$, for $S=1/2$ \cite{Sandvik}.  
This is similar to the
1D system, where the exact solution gives a uniform renormalization
$Z=\pi/2$. 
In 2D however, there are
works which contradict the use of a single uniform renormalization constant. 
Employing the Dyson-Maleev representation of
spin operators, Canali \etal. \cite{Canali} found that the magnon energy at \pihalf is
about 2\% larger than the magnon energy at
\pizero. Expanding around the Ising limit, Singh and
Gelfand \cite{Singh} found a shallow minimum in the dispersion around \pizero
giving the magnon at \pizero about 7\% less energy than at
($\pi/2$,$\pi/2$). In addition, an approach to the Heisenberg
Hamiltonian starting from the $\pi$-flux phase, a state with
short-range antiferromagnetic order, predicted a deep local minimum
around \pizero in the Brillouin zone\cite{Hsu}. 

To clarify this issue we have calculated the dynamic structure
function $S(q,\omega)$ using the quantum Monte Carlo loop
algorithm \cite{Evertz}, which among other useful features operates
in continuous imaginary time \cite{Beard}.
While the simulations are performed in continuous imaginary time, the
measurements of the spin-spin correlation function were written to an
array with typically 200 points in the imaginary time direction. 
Typically
$10^5$ configurations were used both for equilibration and
measurements using a single-cluster implementation of the loop
algorithm, and all data points are averages of at least five
independent runs. 
The focus on high energy peaks makes it sufficient to do measurements
at intermediate temperatures, $T \sim J$, thus avoiding the low
temperature region where the loop algorithm performs poorly.

\begin{figure}
\begin{center}
\epsfig{file=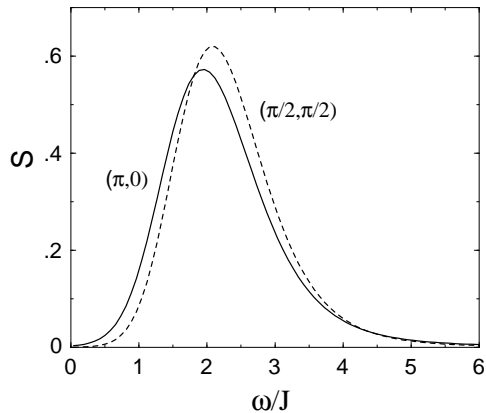,width=6.5cm}
\end{center}
\caption{Dynamic structure function $S(q,\omega)$ at q=\pihalf (dashed line) and q=\pizero (solid line). $T=0.5J$ and $L=32$.
  \label{Sqw}}
\end{figure}

To get real-time dynamics from the imaginary time
data we employed the maximum entropy method \cite{Linden}, with a flat
prior. 
As we restrict ourselves to only calculating {\em relative} magnon
energies at two points in the Brillouin-zone, this choice of
{\em a priori} information should not be crucial, although we expect that
the flat prior overestimates the peak widths. As a check on the
continuation procedure we evaluated the sum rules corresponding to the
-1., 0. and 1. moments of the dynamic structure
function \cite{Hohenberg}. They were all within the error bars of the
quantities, $S(q,i\omega_n = 0)/2$, 
$S(q,\tau = 0)$ and $-\epsilon(2-\cos(q_x)-\cos(q_y))/3$,
respectively, which were extracted 
directly from the imaginary time data ($\epsilon$ is the energy per
site). A typical picture of the dynamic structure factor thus obtained is
shown in Figure \ref{Sqw}.

Following the approach of Makivic and Jarrell \cite{Makivic}, we
determined the magnon energy from the normalized first moment
$\omega_{q}$ of the relaxation function
\be 
 F(q,\omega) =  
2(1-e^{-\beta \omega}) S(q,\omega) \left( \beta \omega \chi(q)\right)^{-1}.
\ee
Because there is nothing that breaks the spin rotational symmetry in
our Monte Carlo calculation this relaxation function is an average
over the transverse and longitudinal relaxation function. This average resembles closely what is measured in neutron scattering experiments.

\begin{figure}
\begin{center}
\epsfig{file=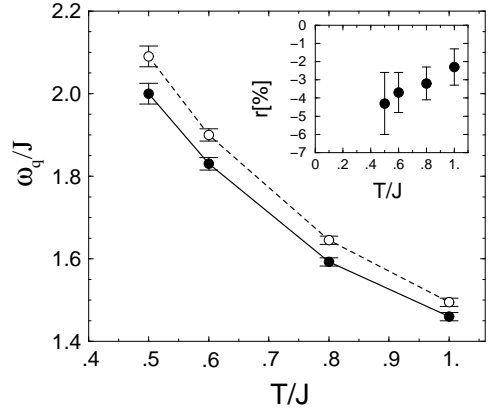,width=6.5cm}
\end{center}
\caption{Magnon energies for the two points \pizero (filled circles)
  and \pihalf (open circles) in the Brillouin zone at different
  temperatures. The inset shows the relative difference $r =
  (\omega_{(\pi,0)}-\omega_{(\pi/2,\pi/2)})/\omega_{(\pi/2,\pi/2)}$ in
  percent. \label{Magnon-energies}}
\end{figure}

Computation of $\omega_q$ for our smallest systems, $4\times 4$, gives
no significant difference between  $\omega_{(\pi,0)}$ and $\omega_{(\pi/2,\pi/2)}$.
This is in agreement with
the exact diagonalization study on small systems of Chen \etal. \cite{Chen}. However,
lattices with $8\times 8$ sites show a clear difference between
$\omega_{(\pi,0)}$ and $\omega_{(\pi/2,\pi/2)}$ for all temperatures
studied. Performing a finite size analysis for $L \times L$-systems,
$L=(4,8,16,32)$, we find that our results are consistent
with the finite size behaviour \cite{Chen}  $\omega_{q,L} \approx
\omega_{q,\infty} + A_q/ L^3$, where $A_q$ is weakly temperature
dependent and of order 10, and $ A_{(\pi/2,\pi/2)} < A_{(\pi,0)}
$. The resulting magnon energies for the infinite size system are
plotted in Figure \ref{Magnon-energies} as functions of temperature. 

It is clearly seen that the magnon energy at \pizero is {\em lower}
than the magnon energy at ($\pi/2$,$\pi/2$). Extrapolating the relative difference to zero temperature 
we find that
the magnon energy at \pizero is about 6\% lower than at
($\pi/2$,$\pi/2$). 
This is in rough agreement with the result obtained by
expanding around the Ising limit \cite{Singh}.

Comparing this result with the measurements, it is seen that 
the zone boundary dispersion in Cu(DCOO)$_2\cdot$4D$_2$O is indeed
accounted for by the correction to spin-wave theory.
Evidence of the zone boundary dispersion has also been observed in
Sr$_2$Cu$_3$O$_4$Cl$_2$ which host two interpenetrating square
lattices of $S=1/2$ spins \cite{kim99}.
La$_2$CuO$_4$ on the other hand show a significant deviation from our 
result (19\%). This is most likely due to higher-order spin couplings in 
the Hamiltonian describing this system.

A qualitative explanation for the zone boundary dispersion can be achieved by
considering the Hubbard model which for large-$U/t$ at 
half-filling is equivalent to the Heisenberg model. A reasonable
ground state ansatz for the Hubbard model which has a low energy and
respects time-reversal invariance is the $\pi$-flux state in which the
electrons behave as if they were subjected to a magnetic field of flux
$\pi$ per plaquette.
In this state the Fermi ``surface'' is located at the points
($\pi/2$,$\pm \pi/2$) in the Brillouin zone, and as the magnons are
particle-hole excitations their dispersion will have minima at
($0$,$\pi$) and ($\pi$,$0$) as well as at (0,0) and
($\pi$,$\pi$). This qualitatively explains the zone boundary
dispersion, but implies at the quantitative level gapless magnons at
($\pi$,0) and (0,$\pi$). However, the $\pi$-flux phase has no
antiferromagnetic long-range order, and so it is necessary to do a
more refined ground state ansatz to get a quantitative explanation. By
considering a ground state consisting of both a 
$\pi$-flux state and a spin density wave (SDW) state which does have a
staggered magnetic moment, Hsu\cite{Hsu} found using a
Gutzwiller projection technique that the total energy of this combined
state could be decreased from the pure $\pi$-flux state.
Using the random phase approximation he found a shallower zone
boundary minimum than for the pure $\pi$-flux state. 
Though still much deeper than what is observed in the present study,
the depth of the zone boundary dispersion at ($\pi$,0) depends crucially
on how much of the ground state is an SDW, 
as a pure SDW state gives a dispersion identical to linear spin wave 
theory\cite{Wen}. It is at present unclear whether a more refined analysis
would lead to a better quantitative agreement with our result, but we feel that 
attributing the zone boundary dispersion to properties of 
the $\pi$-flux state is a plausible, simple and attractive scenario.    

Until now, there has been an unsatisfactory situation, where different
approximative analytical approaches disagreed on the existence of a
zone boundary dispersion in the 2D $S=1/2$ Heisenberg antiferromagnet on a
square lattice. Recent experiments on physical realizations of the
model system have strengthened the need to resolve this question.
By performing finite temperature quantum Monte Carlo simulations, we
have established that indeed there is a zone boundary dispersion. In a
system, which is otherwise well described by a uniform
renormalization of linear spin-wave theory, the zone boundary
dispersion is a remarkable quantum effect.

\ack

We would like to thank Patrick Lee, Radu Coldea {\em et al.} and Rajiv
R. P. Singh for useful discussions, 
and acknowledge the use of the supercomputing facilities at
UNI-C, Aarhus University.


\end{document}